# Ten Conceptual Dimensions of Context [1]

Hashai Papneja [2]

## Abstract

*This paper attempts to synthesize various conceptualizations of the term 'context' as found in computing literature. Ten conceptual dimensions of context thus emerge—location; user, task, and system characteristics; physical, social, organizational, and cultural environments; time-related aspects, and historical information. Together, the ten dimensions of context provide a comprehensive view of the notion of context, and allow for a more systematic examination of the influence of context and contextual information on human-system or human-AI interactions.*

## 1. Introduction

Extant literature on computing offers several definitions of the term *context*. While these definitions are valuable in their own right, this paper attempts to synthesize the key dimensions of context that emerge from these definitions. Whereas prior research has mostly captured technical aspects relating to how context-aware systems have leveraged contextual information (e.g., 1, 2), this study attempts to capture and synthesize the key conceptual dimensions of context. A conceptual synthesis can help inform the design of context-aware applications by identifying pertinent types or categories of contextual information as well as specific attributes within each category that can be leveraged by such applications. It can also inform researchers by helping to identify aspects of context that have perhaps not been fully explored or investigated thus far, but whose influences on pertinent outcomes can be more systematically examined.

## 2. Related Work

Context has been extensively studied in computing literature. Two widely-cited literature reviews on context-aware systems are Baldauf, Dustdar (1) and Hong, Suh (2). Baldauf, Dustdar (1) derive a design framework by examining several context-aware architectures found in literature. Similarly, Hong, Suh (2) review literature on context-aware systems from 2000 to 2007, and propose a new classification framework for such systems based on their technical architectures.

While the above reviews cover technical aspects of context-aware systems and how they capture and manage contextual information, a proper examination of how the concept of context has been defined is lacking. This is also true of numerous other studies on context-aware systems which tend to cover technical aspects of such systems and put forth architectures with a view to effectively implement such systems, but do not squarely focus on how the notion of context has

---

been conceptualized. The purpose of this paper, therefore, is to examine how the notion of context has been defined and conceptualized, identifying the various dimensions of context, and proposing new ones that may also be pertinent to human-system interactions.

## 3. Conceptualizations of Context

In order to understand how context has been conceptualized, we searched for articles containing the terms "context" or "context-aware" in computing literature within the period 1990 to 2010. We attempt to capture the breadth of the discourse on conceptual definitions of context, so studies on context modeling (i.e., how to implement the collection, storage, and dissemination of contextual information) were excluded from the analysis. Eight dimensions of context thus emerged— location, user, task, and system characteristics, physical environment, social environment, time-related aspects, and various historical information. Two additional dimensions were noted from the ISO 9421 standard on human-system interaction—organizational environment, and cultural environment.

Conceptualizations of context can be broadly classified as either representational (including hierarchical) or interactional—terms borrowed from Dourish (3). As Dourish notes, a representational perspective is one where what constitutes as context can be specified in advance, is stable, and is separate from the activity being performed. On the other hand, an interactional perspective is one where context arises from the activity, is an occasioned property of it, and can only be defined dynamically (3). While early conceptualizations of context may have been representational in nature, activity or task information quickly constituted part of later ones. A chronological account of how conceptualizations of context evolved ensues.

From their pioneering work at Xerox PARC, Schilit, Adams (4) identify three aspects of context: "where you are, who you are with, and what resources are nearby" (p. 85), and specify that context encompasses more than just the user's location to also include "lighting, noise level, network connectivity, communication costs, communication bandwidth. and even the social situation: e.g., whether you are with your manager or with a co-worker" (p. 85). Brown (5) offers the following as examples of contextual information: location, adjacency of other objects, critical states, computer states, imaginary companions (which I construe as the user's social environment), and time. Dey (6) defines context as "includ[ing], but is not limited to, information the user is attending to, emotional state, focus of attention, location and orientation, date and time of day, objects and people in the user's environment" (p. 51). Similarly, Schmidt, Beigl (7) propose a hierarchical view of context as one comprising of human factors and the physical environment, whereby human factors comprise of information about the user, the social environment, and the task, while physical environment comprises of information relating to the location, infrastructure, and physical conditions. Building upon Schilit, Adams (4's) and Schmidt, Beigl (7), Gwizdka (8) conceptualizes context as comprising of internal context and external context. Internal context information describes the user's state, and comprises of the user's goals and tasks, work context, personal events, and the user's cognitive, emotional, and physical states. External context information describes the environment's state, and comprises of location, proximity to other objects (both people and devices), temperature, time, etc. Similarly, Lieberman and Selker (9) define context as "everything that affects the computation except the explicit input and output" (p. 618), and conceptualize it as comprising of information relating to the state of the user, the physical environment, the computational environment, and the history of the user-computer-environment interaction. Also taking an interactional perspective of context, Dey, Abowd (10) define it as "any information that can be used to characterize the situation of entities (i.e., whether a person, place, or object) that are considered relevant to the interaction between a user and an application, including the user and the application themselves. Context is typically the location, identity, and state of people, groups, and computational and physical objects" (p. 106), and introduce four categories of context—identity, location, status (or activity), and time. Taking a comprehensive view of context, Prekop and Burnett (11) conceptualize it as having both external (i.e., physical) and internal (i.e., user- and activity-centric) dimensions. Beale and Lonsdale (12) conceptualize context as a hierarchical, dynamic process with historic dependencies. Lastly, the ISO's 9421-210 standard for ergonomics of human-system interaction defines context of use as "the users, tasks, equipment (hardware, software and materials), and the physical and social environments in which a product is used" (13). This definition was later expanded to include organizational and cultural environments.

In summary, conceptualizations of context evolved from restrictive to increasingly comprehensive, and further to also include characteristics of the human-system interaction and activity. Table 1 below lists the dimensions, identifying the articles pertaining to each. The next section synthesizes the discussion relating to each dimension of context in further detail.

| Article | Conceptual Dimensions of Context | | | | | | | | | |
|---|---|---|---|---|---|---|---|---|---|---|
| | Location | Physical Environment | Social Environment | Organizational Environment | Cultural Environment | User Characteristics | System Characteristics | Task- (or Activity-) Specific Characteristics | Time-Related Aspects | Historical Information |
| Schilit, Adams (4) | X | X | X | | | | | | | |
| Brown (5) | X | X | X | | | | X | | X | |
| Dey (6) | X | X | X | | | X | | X | | |
| Schmidt, Beigl (7) | X | X | X | | | X | | X | | X |
| Gwizdka (8) | X | X | X | | | X | | X | | |
| Lieberman and Selker (9) | | X | | | | X | X | X | | X |
| Dey, Abowd (10) | X | X | X | | | X | | X | X | X |
| Prekop and Burnett (11) | X | X | X | | | X | | X | | |
| Beale and Lonsdale (12) | | | | | | X | | X | X | X |
| ISO (13) | | X | X | X | X | X | | X | X | |

Table 1: Conceptual dimensions of context as found in extant literature

# 4. Conceptual Dimensions of Context

The conceptual dimensions of context as found in extant literature on context are as follows:

**Location**
The user's location is the most prototypical conceptualization of context. Location can include the user's spatial location, such as the geographic location (e.g., city, or street, etc.), or a more specific location within a broader location (e.g., meeting room 35 on the 5$^{th}$ floor) (4-6, 8, 11). It can also include the user's orientation and elevation, and information on the relative position and proximity to others (7, 10).

**User Characteristics**
User characteristics encompass information that is directly related to the specific individual using the system. They can refer to the state of the user, including their emotional, cognitive (e.g., current focus-of-attention), and physical (e.g., position) states (6-9, 11, 12). They can also include the user's biophysiological conditions, such as vital signs, or tiredness (7, 10). Lastly, user characteristics can also encompass other information about the user such as their preferences, habits, knowledge, skills, experience, education, and training (7, 13).

**Task Characteristics**
Task characteristics include aspects of the specific task or activity being undertaken by the user. These can include the user's general goals, engaged tasks (e.g., attending a lecture), or even spontaneous activity (e.g., talking, or reading), including their current status (e.g., current projects, their status, and to-do items) (6-13).

**System Characteristics**
System characteristics can refer to the internal state of the computer (e.g., the currently active directory) (5), or the state of the computational environment within which the user is operating (9).

**Physical Environment**
The physical environment can include proximate resources—physical objects such as active badges, printers, displays, speakers, thermostats, furniture, softwares and artifacts such as applications and files, and infrastructure for computation, communication, and task performance—and their attributes such as network connectivity, bandwidth, volume, etc. (4-11, 13). It can also encompass physical attributes of the location, such as the current temperature, ambient light, noise level, etc. (10, 13). Lastly, non-physical resources such as bank accounts, menus, a set of instructions, etc. that can be accessed can also constitute the physical environment (4).

**Social Environment**
The social environment encompasses the people, or imaginary companions (5), in the user's proximal and distal social environments (4, 6, 8, 10, 11, 13), and can also include attributes such as social interactions and group dynamics (7), and characteristics such as group enthusiasm or global mood (10).

**Organizational Environment**
Organizational environment can refer to the relevant stakeholder groups, as well as their relationships with the system described in terms of key goals and constraints (13). It can also include factors such as organizational structure and work practices (13).

**Cultural Environment**
The cultural environment includes factors relating to the overall societal culture such as attitudes of users within the culture (13).

**Time-Related Aspects**
Time-related aspects can include simply the timestamp associated with an artifact or activity (5, 12), the relative ordering of events, a period, an instant, or a time span (10), as well as the way in which users typically carry out tasks—the frequency and duration of performance, interdependencies and activities to be carried out in parallel (13). Time-related aspects are most often used in conjunction with other aspects of context.

**Historical Information**
Historical information mainly comprises of historical values of contextual variables, capturing changes in their values over time to be able to, among other possible reasons, predict their future values (7, 9, 10, 12). For instance, an application may need the user's location history in order to predict their future location (10), or a learner visiting a museum for the second time could have their content recommendations influenced by their activities on a previous visit (12).